# MEASUREMENTS OF NUCLEON-INDUCED FISSION CROSS-SECTIONS OF SEPARATED TUNGSTEN ISOTOPES AND NATURAL TUNGSTEN IN THE 50-200 MEV ENERGY REGION

Eismont V.P.[1], Filatov N.P.[1], Smirnov A.N.[1], Soloviev S.M.[1], Blomgren J.[2], Condé H.[2], Prokofiev A.V.[1,3], Mashnik S.G.[4]

[1] V.G. Khlopin Radium Institute, 2oi Murinskiy Prospect 28, Saint-Petersburg 194021, Russia
[2] Department of Neutron Research, Uppsala University, Box 525, S-751 20 Uppsala, Sweden
[3] The Svedberg Laboratory, Uppsala University, Box 533, S-751 21 Uppsala, Sweden
[4] X-5, Los Alamos National Laboratory, Los Alamos, New Mexico 87545

## Abstract

Neutron- and proton-induced fission cross-sections of separated isotopes of tungsten ($^{182}$W, $^{183}$W, $^{184}$W, and $^{186}$W) and natural tungsten relative to $^{209}$Bi have been measured in the incident nucleon energy region 50 – 200 MeV using fission chambers based on thin-film breakdown counters (TFBC) at quasi-monoenergetic neutrons from the $^{7}$Li(p,n) reaction and at the proton beams of The Svedberg Laboratory (TSL), Uppsala University (Uppsala, Sweden). The preliminary experimental data are presented in comparison with the recent data for nuclei in the lead-bismuth region, as well as with predictions by the CEM03.01 event generator.

## Introduction

Nuclear data, as quantitative characteristics of nuclear reactions, have become particularly important in the intermediate energy range. They are necessary for development of new concepts of nuclear energy production and transmutation of radioactive waste with the use of accelerators, as well as for design of shielding for accelerators and space apparatus, neutron and proton therapy, medical isotope production, and many other applications. They are important also for the development of theory of nuclear interactions, nuclear structure, and nuclear matter properties.

For systems based on high-current accelerators (Accelerator Driven System – ADS) and intended for nuclear energy production and/or transmutation of long-life radioactive waste, these data are necessary first of all for calculations of the neutron production targets. In the report [1], two main options of neutron production targets are considered: fluid (Pb+Bi) eutectic and solid W. For calculations of key parameters of these targets (number and spectrum of emitted neutrons, prompt and residual radioactivity, heat release, radiation stability (in the case of solid target) nuclear data are needed for a number of proton- and neutron-induced reactions, including fission reactions, at intermediate energies. In spite of relatively low cross-section, fission reactions with subactinide nuclei possess high energy release and often lead to products with long half-lives. Fission cross-section of W (isotope $^{184}$W) in the energy region up to the 200 MeV was included in the High Priority Request List [2] due to its importance for development of ADS and nuclear reaction models.

Data for separated tungsten isotopes are of a particular interest for development of nuclear reaction models, because nuclei with atomic weight A≈180 (as $^{182,183,184,186}$W) belong to the region of strongly deformed nuclei (with deformation parameter $\beta_2 \sim 0.25\text{-}0.30$). A comparison of fission cross-sections of these nuclei, and therefore their fissilities, $P_f$, with ones of transitional nuclei with A ≈190-200, for example $^{197}$Au ($\beta_2$~ 0.10-0.15), as well as with ones of the spherical double magic nucleus ($^{208}$Pb) is of interest for development of an accurate model of fission process.

Recently, evaluated cross-section data libraries have been created at LANL for all stable tungsten isotopes up to 150 MeV [3]. However, there are no evaluated data sets for the fission cross-sections, because the experimental database on W has been extremely poor up to now. The early measurements of the $^{nat}$W(n,f) cross-section by Dzhelepov *et al.* [4] were rather of a qualitative character. Recently, the $^{nat}$W(n,f) cross-sections have been measured in the 50 – 180 MeV by the collaboration between the Khlopin Radium Institute and Uppsala University at quasi-monoenergetic neutrons from the $^{7}$Li(p,n) reaction at the neutron beam facility of the TSL [5]. Results of the $^{nat}$W(n,f) cross-section measurement at the "white spectrum" neutron source of the GNEIS facility have been reported in Ref. [6]. Up to now, only one measurement of the $^{nat}$W(p,f) cross-section has been performed in the 50-200 MeV energy region, by Duijvestijn et al. at a single proton energy of 190 MeV [7]. At present, there are no data on proton- and neutron-induced fission cross-sections on separated isotopes of W, except the $^{183}$W(p,f) cross-section measurements at proton energies below 30 MeV [8].

The present work is devoted to measurements and comparison of the neutron- and proton-induced fission cross-sections of $^{182-184,186,nat}$W in the 50-200 MeV energy range. The study is a continuation of a long-term program in the framework of the ISTC projects on fission cross-section measurements for a wide range of heavy nuclei from $^{243}$Am to $^{181}$Ta [5, 9, 10].

**Experimental procedure and processing of the results**

The (n,f) and (p,f) cross-sections for separated isotopes of tungsten ($^{182,183,184,186}$W) and $^{nat}$W relative to the $^{209}$Bi(n,f) and $^{209}$Bi(p,f) cross-sections, respectively, were measured at the quasi-monoenergetic neutron beam facility [11] and at the broad proton beam facility [12] of TSL. Contrary to our previous measurements [5, 9, 10], the neutron part of the present work has been carried out at the upgraded neutron beam facility of TSL [11]. Similar to our previous work [10], the emphasis was put on the similarity of measurement conditions for proton and neutron beams. The proton beam was scattered and then collimated [12] in order to make the beam profile similar to the one for the neutron beam. The same fission chambers based on the TFBC [13] were used at both neutron and proton beams. In order to meet the counting statistics requirements, the neutron measurements have been carried out at a distance of about 1 m from the Li neutron production target, inside the deflecting magnet, at angles of about 1° or 7° with respect to the beam axis. To eliminate background from scattered primary protons, stainless steel absorbers were placed in front of the fission chambers. Threshold properties of the TFBCs allowed separation of fission fragments from any other charged particles and γ-radiation background accompanying the fission process in the neutron and proton beams.

The design of the TFBC fission chambers and the data acquisition system were similar to the ones described in detail in Ref. [5]. The only differences were in the number of detector-target sandwiches (4 instead of 6) and their sensitive area (2.3 cm$^2$ instead of 1 cm$^2$) in the mosaic arrangements. The measurements at neutrons were carried out simultaneously for all separated tungsten isotopes, as well as for the natural element. For measurements with protons, the mosaic arrangements were irradiated in pairs, and one of them was used as a relative monitor of the proton flux. In addition, the proton flux was monitored by detection of protons scattered by a 0.1-mm thick stainless steel exit window of the beam line at an angle of about $45^0$ by means of a scintillation telescope.

The fissile targets were made from metal powders of separated isotopes of tungsten. The metal powders were transformed to the $WO_3$ chemical form and then deposited onto aluminum backings by means of evaporation in vacuum. The thicknesses of the deposited layers, about 2 mg/cm$^2$, have been defined by direct weighting and by method of Rutherford backscattering of α-particles [14].

The measurements of relative counting rates of fission events have been performed at neutron energies of 75, 94, 143, and 174 MeV, as well as at proton energies of 44, 59, 94, 133, and 170 MeV. The total statistical uncertainties of the relative results depend on the projectile energy and amount to not more than 10% for the neutron measurements. For proton measurements, they vary from 5-10% above 90 MeV to 15-25% below 60 MeV.

In the data reduction process, the corrections were introduced to the fission event counting rates, connected with:

1) the TFBC detection efficiency taking into account the charge, mass, and kinetic energy distributions of fission fragments, as well as the influence of angular anisotropy and linear momentum transfer;

2) the shape of the neutron spectrum (for neutron part of the measurements);

3) the background of fission events due to the presence of heavy fissile admixtures in the aluminum backings.

The main aspects of the data processing have been described in details in Refs. [5, 10].

**Results**

Absolute (n,f) and (p,f) cross-sections for separated isotopes of tungsten ($^{182,183,184,186}$W) and $^{nat}$W have been obtained by multiplication of the measured cross-section ratios to the parameterized $^{209}$Bi(n,f) cross-section [5] and the $^{209}$Bi(p,f) cross-section [10,15], respectively. The results are presented in Tables 1 and 2 and shown in Fig. 1.

**Table 1.** Neutron-induced fission cross-sections of the tungsten isotopes.

| $E_n$ (MeV) | $^{182}$W (mb) | Stat. err. (mb) | $^{183}$W (mb) | Stat. err. (mb) | $^{184}$W (mb) | Stat. err. (mb) | $^{186}$W (mb) | Stat. err. (mb) |
|---|---|---|---|---|---|---|---|---|
| 75 | 0.0664 | 0.0048 | 0.0621 | 0.0047 | 0.0371 | 0.0033 | 0.022 | 0.0084 |
| 94 | 0.245 | 0.014 | 0.172 | 0.011 | 0.135 | 0.0094 | 0.0654 | 0.006 |
| 143 | 0.939 | 0.066 | 0.591 | 0.049 | 0.571 | 0.047 | 0.310 | 0.036 |
| 174 | 1.35 | 0.09 | 1.07 | 0.08 | 0.96 | 0.10 | 0.452 | 0.049 |

**Table 2.** Proton-induced fission cross-sections of the tungsten isotopes.

| $E_p$ (MeV) | $^{182}$W(p,f) (mb) | Stat. err. (mb) | $^{183}$W(p,f) (mb) | Stat. err. (mb) | $^{184}$W(p,f) (mb) | Stat. err. (mb) | $^{186}$W(p,f) (mb) | Stat. err. (mb) |
|---|---|---|---|---|---|---|---|---|
| 44.3 | 0.0088 | 0.0017 | 0.0082 | 0.0022 | 0.00564 | 0.0015 | 0.0045 | 0.0011 |
| 59.9 | 0.065 | 0.010 | 0.059 | 0.010 | 0.0277 | 0.0049 | 0.0196 | 0.0034 |
| 93.5 | 1.24 | 0.06 | 0.872 | 0.046 | 0.595 | 0.031 | 0.325 | 0.020 |
| 132.5 | 2.74 | 0.18 | 2.11 | 0.14 | 1.73 | 0.12 | 0.953 | 0.061 |
| 169.6 | 4.37 | 0.25 | 3.46 | 0.19 | 2.88 | 0.16 | 1.80 | 0.10 |

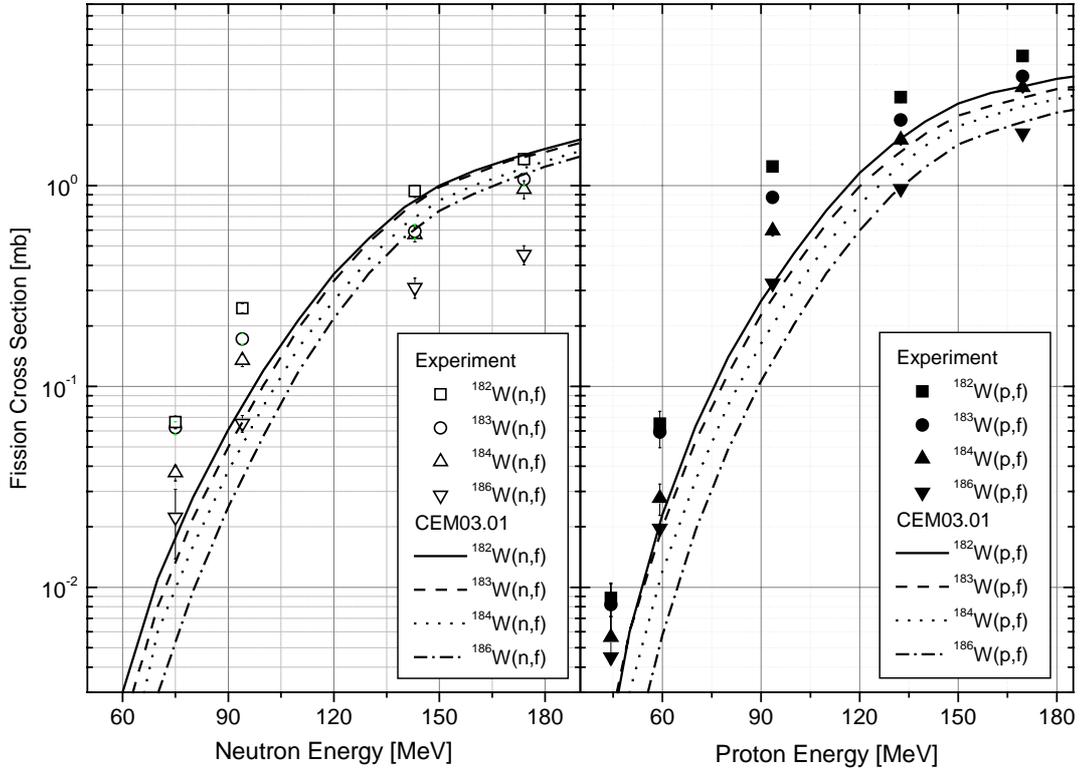

FIG. 1. The (n,f) and (p,f) cross-sections of $^{182}$W, $^{183}$W, $^{184}$W, and $^{186}$W versus incident nucleon energy. Symbols are experimental results of the present work, and curves are results of calculations by the CEM03.01 event generator [16].

In Fig. 1, the experimental data are compared with the results of calculations by the CEM03.01 event generator [16]. It is seen that CEM03.01 describes the data reasonably well at energies above about 130 MeV (except of the data for $^{186}$W(n,f)), but underestimates both the proton- and neutron-induced fission cross-sections at lower energies for all measured tungsten isotopes. This situation may be understood taking into account the deformation of tungsten nuclei (contrary to lead nuclei), i.e. competition of the neutron channel can reduce the cross-section growth due to rotational enhancement of the level density [17].

In view of practical importance of the data for $^{nat}$W, we present also the (n,f) and (p,f) cross-section dependences on incident nucleon energy obtained both by direct measurements using the $^{nat}$W samples and by calculations on a basis of results for separated isotope samples taking into account their contributions in the natural element. The results of the present work are given in Table 3 and in Fig. 2 (together with our earlier results on $^{nat}$W(n,f) [5] and the result of Duijvestijn *et al.* on $^{nat}$W(p,f) at 190 MeV [7]. It is seen from the figure that all data are in good agreement.

Since the (n,f) and (p,f) cross-sections were measured at different projectile energies, the absolute experimental data were fitted by the exponential functions [5] in order to provide a more convenient basis for further comparative analysis of the cross-sections.

**Table 3.** Neutron- and proton-induced fission cross-sections of natural tungsten. The calculated results were obtained by summation of the measured data for separated tungsten isotopes (see Tables 1,2) taking into account their contributions in the natural element.

| $E_n$ (MeV) | $^{nat}$W(n,f) meas. (mb) | Stat. err. (mb) | $^{nat}$W(n,f) calc. (mb) | Stat. err. (mb) | $E_p$ (MeV) | $^{nat}$W(p,f) meas. (mb) | Stat. err. (mb) | $^{nat}$W(p,f) calc. (mb) | Stat. err. (mb) |
|---|---|---|---|---|---|---|---|---|---|
|  |  |  |  |  | 44.3 |  |  | 0.0065 | 0.0015 |
| 75 | 0.040 | 0.006 | 0.044 | 0.005 | 59.9 | 0.035 | 0.006 | 0.040 | 0.007 |
| 94 | 0.139 | 0.010 | 0.149 | 0.010 | 93.5 | 0.655 | 0.037 | 0.727 | 0.038 |
| 143 | 0.577 | 0.048 | 0.595 | 0.049 | 132.5 | 1.7 | 0.1 | 1.83 | 0.12 |
| 174 | 0.95 | 0.08 | 0.93 | 0.08 | 169.6 | 3.0 | 0.2 | 3.04 | 0.17 |

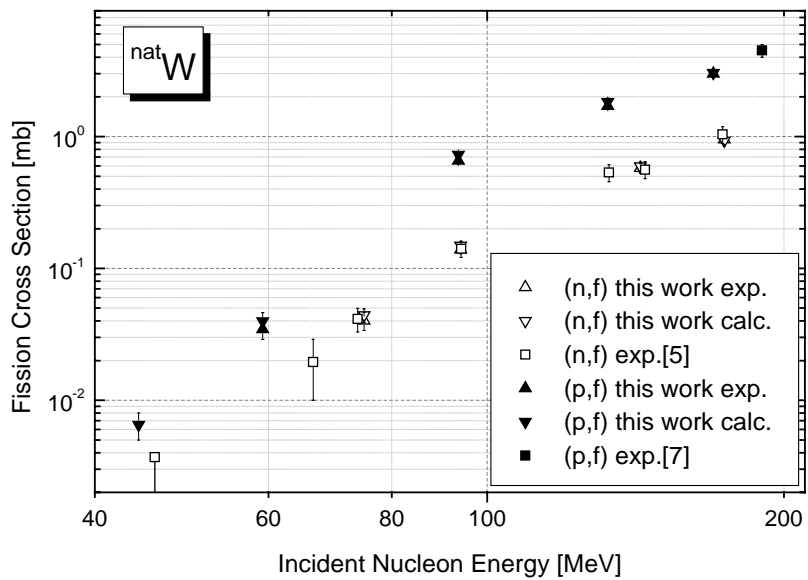

FIG. 4. The (n,f) and (p,f) cross-sections of $^{nat}$W versus incident nucleon energy.

**Discussion**

The fission cross-section, $\sigma_f$, is determined by the reaction cross-section $\sigma_r$, and the fission probability of the created nucleus, $P_f$, i.e., $\sigma_f = \sigma_r P_f$. Therefore a study of the influence of isotope spin of the incident particle and the target nuclei allows better understanding both the mechanism of nucleon-nucleus interaction (via the properties of the optical potential) and the mechanism of fission of excited nuclei. In the present paper the dependence is examined of measured values of fission cross-sections on the fissility parameter, $Z^2/A$, as characteristics of the second stage of the interaction – the probability of decay (fission) of a nucleus.

At present, isotopes of tungsten are the lightest nuclei for which the study of influence of isotope effects in a wide range of excitation energy has been carried out. A low contribution of the shell correction (~1 MeV) to the fission barrier (~30 MeV) and a large deformation ($\beta \approx 0.25$) for the tungsten isotopes distinguish them significantly from the lead isotopes studied earlier. The latter are of spherical shape and have the largest microscopic contribution to the barrier of all nuclei (~15 MeV). Thus, the tungsten isotopes can be considered as representatives of the typical liquid-drop fission.

First of all, the dependence should be mentioned of the (p,f)/(n,f) cross-section ratio on the parameter $Z^2/A$ (of a target nuclei in this case). This dependence was first studied in Ref. [18], and this has led to a large series of measurements for different nuclei. It has been observed that the $\sigma_{pf}/\sigma_{nf}$ ratio depends strongly on the incident nucleon energy in the low-energy region (20 – 70 MeV) and then approaches a plateau slowly. In Fig. 3, results for the incident nucleon energy 180 MeV are shown. It is seen that the $\sigma_{pf}/\sigma_{nf}$ ratio increases with the $Z^2/A$ parameter decrease. However, it is also seen that the $\sigma_{pf}/\sigma_{nf}$ ratio growth is not monotonous, but it is slowed down (or stopped) in case of transition from the lead isotopes group to the tungsten one. This effect has been predicted earlier in the work [19] and was ascribed to connection of the fission cross-section ratio with a value of the fission barrier. The fission barrier changes weakly for heavy actinide nuclei. In the region of lighter nuclei the barrier increases rapidly with the nuclear mass and charge decrease and reaches a maximum value for $^{208}$Pb due to large value of the shell correction to the liquid-drop barrier. Since the shell correction decreases, the barrier reaches a plateau that can lead to slowing down of the $\sigma_{pf}/\sigma_{nf}$ rise in the region of tungsten.

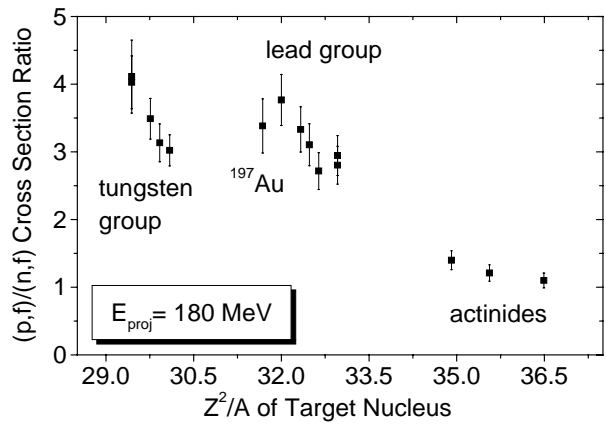

**Fig. 3.** The (p,f)/(n,f) cross-section ratios versus $Z^2/A$.

Differences in the nature of the barriers (relation between microscopic and macroscopic components) and degrees of deformation of the tungsten and lead isotopes could lead to difference in the dependence of the fission cross-section of these nuclei on the $Z^2/A$ parameter of composite system (incident nucleon + target nucleus). First, it should be due to different dependences of the shell and liquid-drop components of the barrier on temperature (shell correction decreases rapidly with the temperature increase). Second, the deformation of the ground state leads to collective (rotational) enhancement of the level density of the nucleus created after emission of neutrons and thus to increase of the neutron width and repression of the fission cross-section.

However, as it is seen in Fig. 4, in which the dependences of the fission cross-sections on the $Z^2/A$ parameter of the composite system is presented, fission cross-sections for both groups of isotopes at all incident nucleon energies (maybe except the lowest energy) are normally distributed along the same straight lines in a semi-logarithmic scale. It is hard to conclude whether it is the result of damping of shell effects (in the case of lead) at the temperature of fissioning nuclei corresponding to the nucleon energy above 70 MeV ($T \approx \sqrt{\frac{E^*}{a}} \approx \sqrt{\frac{60}{25}} \approx 1.5$ MeV, where $a = A/8$), or the result of change of nucleon composition of the actually fissioning nuclei and their mass, charge and excitation energy distributions in comparison with the composite system. It is also possible that a high fission barrier of doubly magical lead isotopes and neighboring nuclei, leading to decrease of the fission probability, is compensated for by the low level density in the ground state region of spherical nuclei, leading to a decrease of probability of neutron emission [19, 20].

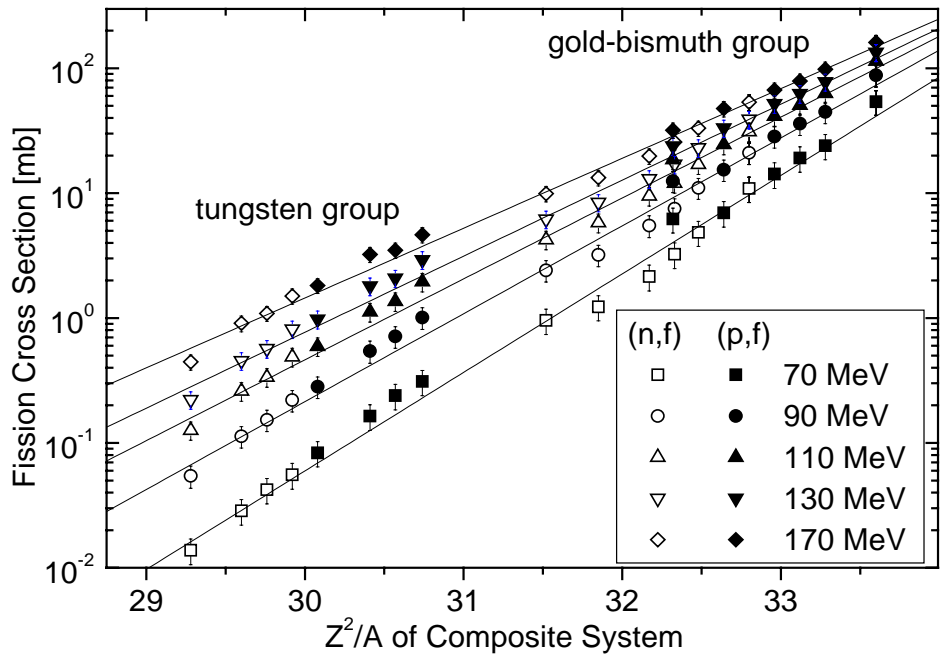

**FIG. 7.** The (n,f) and (p,f) cross-sections of sub-actinide nuclei versus parameter $Z^2/A$ of the composite system for the different incident nucleon energies. The lines are the best fits.

**Conclusion**

The neutron- and proton-induced fission cross-sections of separated isotopes of tungsten ($^{182}$W, $^{183}$W, $^{184}$W, and $^{186}$W) are measured for the first time in the incident nucleon energy region 50 – 200 MeV. The obtained results on (n,f) and (p,f) cross-sections are compared with each other, with the recent data for nuclei in the lead-bismuth region, as well as with predictions by the CEM03.01 event generator. As for the current version of CEM03.01, its underestimation of both proton- and neutron-induced experimental fission cross-sections on tungsten at energies below about 130 MeV observed here is an indication that the determination of the level-density ratio done in Ref. [16] should be redone, ensuring that the latest version of CEM03.01 describes as well as possible fission cross-sections from various reactions.

The work was performed in the framework of the ISTC project 2213 and partially supported by the Advanced Simulation Computing (ASC) Program at the Los Alamos National Laboratory operated by the University of California for the U.S. Department of Energy.